\RequirePackage{fix-cm}
\documentclass[smallextended]{svjour3}       
\smartqed  
\usepackage{graphicx}
\graphicspath{ {figs/} {pics/}}
\usepackage{subcaption}
\usepackage{floatrow}
\usepackage{xcolor}
\usepackage{flafter}
\usepackage{makecell}
\usepackage[final]{changes}

\def\makeheadbox{{%
\hbox to0pt{\vbox{\baselineskip=10dd\hrule\hbox
to\hsize{\vrule\kern3pt\vbox{\kern3pt
\hbox{Pre-print manuscript}
\kern3pt}\hfil\kern3pt\vrule}\hrule}%
\hss}}}

\makeatletter
\let\NAT@parse\undefined
\makeatother
\usepackage{hyperref}  

 \journalname{Pre-print}

\begin{document}

\title{Community membership consistency \replaced{applied to}{in} corporate board interlock networks
}


\author{Dafne E. van Kuppevelt$^1$        \and
        Rena Bakhshi$^1$ \and
        Eelke M. Heemskerk$^2$ \and
        Frank W. Takes$^{2,3}$  
}

\authorrunning{Dafne E. van Kuppevelt        \and
        Rena Bakhshi \and
        Eelke M. Heemskerk \and
        Frank W. Takes }
        
\institute{$^1$ Netherlands eScience Center, Amsterdam, Netherlands \\
           $^2$ Universiteit van Amsterdam, Amsterdam, Netherlands \\
           $^3$ Leiden University, Leiden, Netherlands
}

\date{October 2021\footnote{This manuscript has been accepted for publication at the Journal of Computational Social Science}}

\maketitle

\begin{abstract}
Community detection is a well established method for studying the meso scale structure of social networks. 
Applying a community detection algorithm results in a division of a network into communities that is often used to inspect and reason about community membership of specific nodes. 
This micro level interpretation step of community structure is a crucial step in typical social science research.
However, the methodological caveat in this step is that virtually all modern community detection methods are non-deterministic and based on randomization and approximated results. 
This needs to be explicitly taken into consideration when reasoning about community membership of individual nodes. 
To do so, we propose a metric of \emph{community membership consistency}, that provides node-level insights in how reliable the placement of that node into a community really is. 
In addition, it enables us to distinguish the \emph{community core} members of a community.
The usefulness \added{of} the proposed metrics is demonstrated on corporate board interlock networks, in which weighted links represent shared senior level directors between firms. 
Results suggest that the community structure of global business groups is centered around persistent communities consisting of core countries tied by geographical and cultural proximity. 
In addition, we identify fringe countries that appear to associate with a number of different global business communities. 
\end{abstract}


\newpage

\section{Introduction}
\label{sec:introduction}

Community detection has established itself as method for detecting groups in social systems, unveiling the meso level structure of networked environments. 
The obtained division of a network in communities is often used in the social sciences to understand how individual nodes in the network belong to particular communities and how strong or durable such an affiliation is. 

Unfortunately, a general problem with community detection methods is the inherent uncertainty as a result of randomization and approximation. 
This hinders interpretation of community membership on the node level. 
Therefore, we propose a solution to this methodological challenge and demonstrate how this leads to meaningful insights. In order to illustrate our approach we consider networks of interlocking directorates, where weighted ties represent shared directors between firms. 
These so-called board interlock networks have been extensively studied in corporate governance and social network analysis literature~\cite{carroll2013making,kogut2012small,davis2003small,davis1991agents,carroll2004corporate,takes2016centrality,valeeva2020duality}. 
The community structure of these board interlock networks has been shown to provide valuable insights related to global business elites and transnationalization~\cite{heemskerk2016corporate,Heemskerk2016WhereDirectorates}.

Multiple community detection methods and algorithms have been developed over the past decades, each returning a different division of the network into communities~\cite{Blondel2008FastNetworks,traag2019leiden,rosvall2008maps,Lancichinetti_Radicchi_Ramasco_Fortunato_2011}. 
Several methods are based on the notion of optimizing a quality score, such as modularity \cite{newman2004modularity}. 
This score indicates for a given network divided into communities, also called a clustering, how well this was done; often in some way favoring many links within a community, and few links between communities. This is in line with theoretical as well as intuitive understandings of the concept of community. As this process of optimization is analytically intractable, popular algorithms available in standard network analysis tooling, such as Louvain \cite{Blondel2008FastNetworks} and the Leiden algorithm \cite{traag2019leiden}, use heuristics to find a high-quality clustering. 
Typically, these heuristic algorithms, upon multiple runs, return different divisions of the network into communities. 
These solutions may all be optimal or near-optimal solutions with a high quality score. 
However, the solutions may differ substantially in terms of which node is in which community. 
This is in some contexts referred to as the degeneracy of multiple solutions~\cite{Good2010PerformanceContexts}. 
While this is a direct result of the randomness and heuristics involved in the underlying algorithms, it hinders scholars from meaningfully interpreting community detection results, especially when this interpretation takes place at the level of individual nodes. 

A method that has been proposed to solve this problem, is consensus clustering \cite{Lancichinetti_Fortunato_2012}. In this method, multiple runs of a community detection algorithm help determine the most consistent clustering of the network into communities. 
This approach mitigates the uncertainty of one community detection solution, obtaining in the clustering for which there is the most `consensus' across many runs of the algorithm. \added{An accelerated version of the algorithm was introduced in \cite{tandon2019fast}.}
While interesting for providing a stable meso level view of the network, there may still be substantial differences in terms of which nodes are consistently placed in the same community across multiple runs of the algorithm. Regardless, the final outcome of the consensus clustering algorithm is an assignment of all nodes to a particular community, in essence treating each node as being an equal member of its community. This makes it challenging to reason about individual nodes and their community membership, as we do not know whether this particular node was consistently in the same community. However, it is exactly this type of community membership information that is often needed to obtain actionable insights in real-world networked systems~\cite{nagy2018friendship,sporns_structure_2013,de_leo_community_2013}, including the board interlock networks considered in this paper, as we will discuss later.

The problem of proper interpretation of the community detection outcomes becomes even more pronounced when scholars move to a comparative approach and compare different community detection solutions. 
After all, the differences such a comparison uncovers may very well be because particular nodes did not ``fit'' very well in one particular community. 
Examples of such comparative studies include replications, but also longitudinal studies in which we ultimately want to understand whether a node that moves from one to another community is actually doing so as a result of substantive change \deleted{d} in the underlying system. 

The  discrepancy between the technical solution to the problem of community consensus at the meso level and reliable community membership inference at the micro level, is the topic of this study. 
The latter is crucial for proper interpretation of the community detection results. 
To close this gap, we propose a metric for nodes called \emph{community membership consistency}. 
This metric assesses the extent to which a node's community assignment is consistent across different runs of the community detection algorithm. 
Ultimately, in addition to the actual node's community membership based on consensus clustering, it allows one to quantify how consistent this node's community membership actually is.  
Moreover, the consistency score can be used to distinguish between \emph{community core and fringe} members. 

Several related consistency metrics have been proposed in previous works, but often with a goal different than interpreting an individual node's community membership. 
For example, these are metrics with the aim of defining new centrality measures based on the consensus between different runs of a community detection algorithm~\cite{kim_community_2015,kim_relational_2019}, or with the goal of proposing a method to find more stable communities \cite{seifi_stable_2013}. 
The concept of \added{community} cores or `building blocks' has been explored in previous work as well. 
In \cite{seifi_stable_2013}, the consensus matrix is used to define cores that are (almost) always placed in the same community. 
The authors of \cite{Riolo_Newman_2019} take an information-theoretic approach, optimizing for building blocks that maximizes the mutual information of community assignment, conditioned on the building blocks. 
A similar study looked at invariant groups of nodes in communities and investigated their properties~\cite{Chakraborty_Srinivasan_Ganguly_Bhowmick_Mukherjee_2013}. In \cite{gilbert2018clusters}, the outcomes of a number of runs of community detection are used for semi-supervised learning, expanding a seed set of nodes based on a similarity measure computed from the consensus matrix. 
A recent study \cite{Peixoto_2020} proposes a method for computing the consensus and dissensus between the degenerate partitions, by aligning the different clusterings and describing the posterior distribution based on a stochastic block model.

Compared to the works discussed above, our approach differs in the sense that it focuses on enabling the interpretation of results in an actual computational social science context. 
This means that we aim for interpretable metrics that provide an understanding of community membership at the level of an individual node, as well as community core formation of multiple nodes.

In this paper, we use \deleted{the proposed metric to analyze}\added{ board interlock networks as an illustrative example for the proposed metric}. 
Board interlocks, where two firms share at least one board member, are widely studied in order to understand the network structure in corporate governance and corporate elites. 
These corporate networks allow scholars to investigate how corporations and the individuals involved exert power over others, gain access to information and in general interact within the global economy.  
Exploring corporate network structures using network analysis techniques has greatly improved our understanding of the global corporate system~\cite{kogut2012small,davis2003small,takes2016centrality,heemskerk2016corporate}.
Network studies have aided in unraveling the spread of corporate practices~\cite{davis1991agents}, the formation of a corporate elite~\cite{davis2003small,carroll2004corporate}, and the formation of business groups and elite transnationalization~\cite{heemskerk2016corporate}.
We extend on this line of research by studying the consistency of the community structure of the board interlock network at the node level. 
We will investigate and compare the community detection results at three levels of granularity: at the level of firms, at the aggregated level of cities, and at the aggregated level of countries, similar to how this is done in previous work~\cite{takes2016centrality,Heemskerk2016WhereDirectorates,heemskerk2016corporate}. 
We focus on the last level of aggregation when zooming in on individual nodes in the network, describing their community membership in relation to their consistency score. 
For this particular network, we look in detail at community cores and fringes, allowing us to assess which countries form consistent clusters of power, and which countries are at the fringe of these power centers. 

The rest of the paper is structured as follows. In Section~\ref{sec:method} we explain our methodology and the concepts needed to define the proposed metrics of community membership consistency and community cores. 
In Section~\ref{sec:results} we present our results of applying these measures to board interlock network data, both on a high level for all three networks, and in a more detailed manner for the country level network. Section~\ref{sec:discussion} discusses the implications of the results for the research field as well as directions for future work, whereas finally Section \ref{sec:conclusion} concludes the paper and summarizes the main findings.

\section{Method} \label{sec:method}
In this section we describe our methodology, of which the input is a network, and the output is a division into communities together with interpretable node consistency scores, that can subsequently be used for identifying community cores. 
The approach builds on consensus clustering (Section~\ref{sec:consensus}), after which a number of edge- and node-specific measures are derived in Section~\ref{sec:consistencyscores} in order to compute the proposed measure of community membership consistency in Section~\ref{sec:consistencymeasure}. 
Finally, the approach to \deleted{then} derive community cores is explained in Section~\ref{sec:communitycores}. 

\subsection{Consensus clustering}
\label{sec:consensus}

As discussed in Section~\ref{sec:introduction}, a key problem in modern community detection methods is the instability of results across multiple runs of a particular community detection algorithm. 
To overcome this problem and obtain a stable partitioning, we apply consensus clustering~\cite{Lancichinetti_Fortunato_2012} to the input network. 
In this method, a community detection algorithm is applied $n$ times to a network represented as a weighted adjacency matrix $A$, but with different initial random states. 
This means that we obtain multiple clusterings of the network from which a so-called `consensus matrix' can be constructed. 
For each combination of nodes $i$ and $j$, the value of the consensus matrix $c_{ij}$ denotes the fraction of clusterings in which $i$ and $j$ are placed in the same community. 
As a next step, all values in the consensus matrix below a threshold $\tau$ are set to zero. 
Then the filtered consensus matrix is regarded as a weight matrix of a new network, which is used as input to a next iteration of community detection and thresholding. 
This process is repeated until convergence of the clustering, that is, the clustering does not change in subsequent iterations. 
It was found empirically that in most cases, only one to three iterations are needed to obtain convergence. 

Because the method above works \emph{independent of the community detection algorithm} that is used, it can easily be extended to more complex network data. 
While we work with undirected weighted networks in the remainder of this paper, including for example directionality or a multilayer structure is possible as long as the employed community detection algorithm properly handles these aspects, \added{for example using the methodology proposed in \cite{mucha2010community}}. 

In this work we \replaced{exemplify our approach using}{use} the Leiden algorithm \cite{traag2019leiden} for community detection. 
The Leiden algorithm optimizes the modularity score, similar to the well known Louvain algorithm \cite{Blondel2008FastNetworks}. Modularity is defined as:
\[Q = \frac{1}{2m} \sum_{ij}(A_{ij} - \gamma \frac{k_i k_j}{2m})\delta(C_i, C_j)\]
Here, $A_{ij}$ is the adjacency matrix, $m_c$ is the number of edges in community $c$, $C_i$ is the community assignment of node $i$ and $k_i$ is the degree of node $i$. 
The parameter $\gamma$ controls the resolution at which communities are detected, which we leave untouched at a value of $1$.
Whereas the Louvain algorithm heuristically optimizes the modularity score by merging clusters and moving nodes, the Leiden algorithm also includes improvements for overcoming the problem of badly connected clusters that may result from the original Louvain algorithm.

\begin{figure}[!b]
    \centering
    \begin{subfigure}[b]{0.3\textwidth}
        \includegraphics[width=\textwidth]{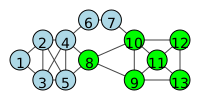}
        \caption{$p$: 5\%, $Q$: 0.37}
    \end{subfigure}
    \begin{subfigure}[b]{0.3\textwidth}
        \includegraphics[width=\textwidth]{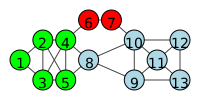}
        \caption{$p$: 29\%, $Q$: 0.40}
        \label{fig:example:b}
    \end{subfigure}
    \begin{subfigure}[b]{0.3\textwidth}
        \includegraphics[width=\textwidth]{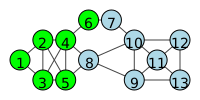}
        \caption{$p$: 3\%, $Q$: 0.37}
    \end{subfigure}
    \begin{subfigure}[b]{0.3\textwidth}
        \includegraphics[width=\textwidth]{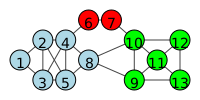}
        \caption{$p$: 57\%, $Q$: 0.40}
        \label{fig:example:d}
    \end{subfigure}
    \begin{subfigure}[b]{0.3\textwidth}
        \includegraphics[width=\textwidth]{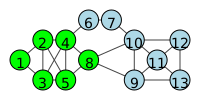}
        \caption{$p$: 6\%, $Q$: 0.37}
    \end{subfigure}
    \begin{subfigure}[b]{0.3\textwidth}
        \includegraphics[width=\textwidth]{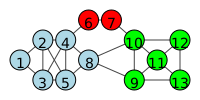}
        \caption{Consensus clustering}
    \end{subfigure}
    \caption{Five divisions into communities (a)--(e) of a toy network (from \cite{kim_relational_2019}), listing $p$, the percentage of iterations resulting in this clustering over 100 runs and corresponding modularity quality score $Q$. The consensus clustering in (f).
    }
    \label{fig:example}
\end{figure}

As an example, Figure \ref{fig:example} shows the result of running the Leiden algorithm for 100 iterations. 
It finds five different clusterings in these 100 runs (Figure \ref{fig:example}(a)--(e)), with modularity values $Q$ that are very close (or equal) to each other.
The value of $p$ indicates the percentage of runs resulting in that particular clustering. 
For this example, the consensus clustering happens to coincide with the most common clustering (Figure \ref{fig:example:d}). Although this is not the case in general, in \cite{Lancichinetti_Fortunato_2012} as well as in our experiments in Section~\ref{sec:results} it is empirically shown that with a proper choice of a threshold value, the quality of the consensus clustering is similar to the individual runs of community detection, but has a more stable character. 
Therefore in the remainder of this work, we choose the threshold value $\tau$ that corresponds to the division of the network into communities with the largest modularity value.

\added{Note that the algorithmic runtime and memory usage of the consensus clustering is also the determining factor in our methodology. 
This can, in theory, get close to quadratic in the number of nodes, because the full consensus matrix is calculated. In subsequent steps, we will only need the elements from the consensus matrix that correspond to edges in the network. Therefore, the fast consensus clustering method from \cite{tandon2019fast} can be used, which uses a reduced consensus matrix, making the complexity of consensus clusering and thus our method linear in the number of edges.}

\subsection{Edge consistency}
\label{sec:consistencyscores}

Consensus clustering results in a stable partitioning, but no insight in how reliable this partitioning  really is.
We therefore take a closer look at the consensus matrix, as calculated in the first iteration of the consensus clustering algorithm. 
For each combination of nodes $i$ and $j$, the consensus value $c_{ij}$ is a value between 0 and 1 that denotes how often those two nodes were clustered together. 
This allows us to define the \textit{node pair consistency} $s_{ij}$ for each node pair $i$,$j$ as follows:
\[s_{ij} = 2\vert c_{ij}-0.5 \vert\]
This consistency value can be considered a dispersion metric of the consensus values, equivalent to the mean absolute difference. The value is multiplied by 2 to scale the consistency value to a more easily interpretable range between $0$ and $1$. 
If $s_{ij}$ is equal to maximum value 1, then this means that the combination of nodes always lies either within one community or in two separate communities, i.e., those two nodes are very consistently placed with respect to each other.  
A value of 0 denotes maximum disagreement between the clusterings: half of the clusterings groups the nodes together and the other half assigns the two nodes to different communities.
Note that the consistency score above is defined for all possible node pairs, including those that are not connected by an edge in the network. 
If nodes pairs are connected with an edge, we will henceforward refer to the corresponding consistency score as the \textit{edge consistency}.

\subsection{Community membership consistency}
\label{sec:consistencymeasure}

Next, we must move from the measure of edge consistency discussed above to a node-specific consistency score. 
Note that this step from an edge-centered metric to a node-centered metric is not trivial. 
We propose to consider the distribution of edge consistency scores for the edges attached to this node, to assess whether this node is consistently put in the same community as its neighbors. 
Between different clusterings, a node might move to a different community together with some of its neighbors. 
This means that the consistency of edges to those neighbors will be high, whereas it will be low for the edges to neighbors that do not move. Recall that our objective is to develop a metric for community membership consistency for each node in our network.  
We do not want to assign a high consistency value to a node that moves together with its neighbors, even though it has some high consistency edges. 
This disqualifies the use of the mean of edge consistency scores over all edges attached to a node, as suitable node-specific score. 
As we will see later in experiments, the distribution of edge consistency values is very skewed, so that the mean value is usually high, while there are clearly some outliers with a very low value. 

To illustrate, see node 7 in the example network of Figure~\ref{fig:example}. Node 7 is almost always clustered together with node 6, so if we would take the mean of edge consistencies from node 7, it would be pulled up by this highly consistent edge. Intuitively, however, we would not assign node 7 a high consistency because, between the different clusterings, it moves communities together with node 6.

As we are interested in the effect of the
low edge consistencies on the node's consistency, we introduce a threshold method. 
For a node $i$ and a chosen threshold \(\theta\) we define the \textit{community membership consistency} $s_i^{\theta}$ of a node as:

\[ s_i^{\theta} = \frac{\sum_j \delta_{ij} \mathbf{1}_{s_{ij}\geq\theta} }{\sum_j  \delta_{ij}} \]

\begin{figure}[!bt]
    \centering
    \begin{subfigure}[b]{0.3\textwidth}
        \includegraphics[width=\textwidth]{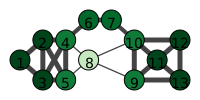}
        \caption{Mean consistency}
    \end{subfigure}
    \begin{subfigure}[b]{0.3\textwidth}
        \includegraphics[width=\textwidth]{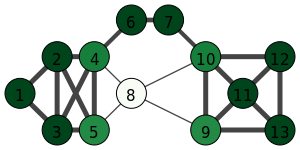}
        \caption{Community membership consistency~$s_i^{0.82}$}
    \end{subfigure}
    \begin{subfigure}[b]{0.3\textwidth}
        \includegraphics[width=\textwidth]{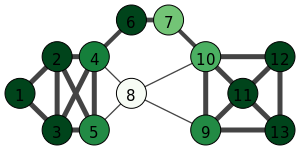}
        \caption{Community membership consistency~$s_i^{0.84}$}
    \end{subfigure}
    \begin{subfigure}[b]{0.3\textwidth}
        \includegraphics[width=\textwidth]{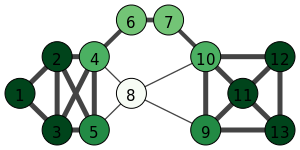}
        \caption{Community membership consistency~$s_i^{0.94}$}
    \end{subfigure}
    \begin{subfigure}[b]{0.3\textwidth}
        \includegraphics[width=\textwidth]{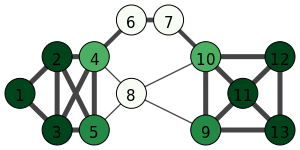}
        \caption{Community membership consistency~$s_i^{1}$}
    \end{subfigure}
    \caption{Community membership consistency, depicted by node color intensity, for \replaced{all possible}{different} thresholds in an example network. Edge consistency, depicted by edge thickness.}
    \label{fig:example-consistency}
\end{figure}

Here, \(\delta_{ij}\) denotes the existence of an edge between node $i$ and $j$, i.e., whether $A_{ij} > 0$.
The community membership consistency $s_i^{\theta}$ essentially denotes the fraction of the edges connected to the node with an edge-consistency larger than \(\theta\). 
Thus, the threshold parameter \(\theta\) denotes how often nodes should be clustered consistently so that we find it trustworthy enough to derive conclusions from it. 
This may depend on the number of partitions we find in the first place, and the modularity landscape they form. To choose a meaningful value for the threshold, we consider \replaced{possible values that follow from the distribution of edge consistencies. Ideally, we choose}{the distribution of community membership consistency for different threshold values and choose} a threshold that is high enough to distinguish inconsistent nodes, but not too high as this may lead to many inconsistent nodes due to small disturbances in the community assignments. 
In general, a higher threshold leads to lower values of node \replaced{consistency}{inconsistency}. 
As an example, see Figure \ref{fig:example-consistency} where community membership consistency is plotted for all possible threshold values in this toy example (see Figure~\ref{fig:example}). \added{The different threshold values in this small example correspond to the unique values of edge consistency in the network.}
Compared to the mean consistency over the node's edges, community membership consistency has a less skewed distribution. We observe the same difference in larger networks, as we will see in Section~\ref{sec:results}.  

\subsection{Community cores}
\label{sec:communitycores}

Similar to how this is done in \cite{seifi_stable_2013}, we can use the community membership consistency scores to reason about groups of nodes that are consistent members of a community.
This allows us to define \replaced{\textit{community cores}}{community \textit{cores}}, that is, set of nodes that are consistently placed together in the same community. 
If a node is almost always placed in the same community as all of its neighbors, it will have a consistency value of exactly $1$. We call these nodes the \textit{hard \added{community} core}. Note that the hard community core could be disconnected, as groups of nodes can be moved to a different community together. 

It is also useful to distinguish in a less strict way, \emph{\added{community} core nodes} that have a high community membership consistency. For this, we choose a threshold \added{$\kappa$} close to 1. Similarly, \emph{fringe} members are nodes that have a very low community membership consistency, for which we choose another\added{, typically much lower,} threshold \added{$\phi$}.
The values of these thresholds can be chosen empirically, possibly considering the robustness of the \added{community} cores around these threshold values.

\section{Experiments}\label{sec:results}

We start by describing the board interlock network data used to evaluate the method in Section~\ref{sec:data}, after which the experimental setup is discussed in Section~\ref{sec:exp_setup}. 
We inspect the results of the consensus clustering in Section~\ref{sec:mod_landscape}, before evaluating the node consistency scores in Section~\ref{sec:node_consistency}. 
An interpretation of the results is given in Section~\ref{sec:descriptive}.

\subsection{Data} \label{sec:data}

We apply the method on board interlock data derived from a 2015 snapshot of the ORBIS database \cite{ORBIS}. 
This database contains global firm-level information, including positions of directors at these firms as well as the firm's location city, country and operating revenue. We use the latter as an indicator of company size, and consider only firms above a certain threshold revenue, as a previous study has shown that the quality of this data is high for large firms \cite{garcia2018effects}.  
From this database, we created the following three network datasets by projecting the raw data on positions of directors at firms to a firm-by-firm board interlock network:

\begin{enumerate}
    \item \textbf{Firm network:} all board interlocks between firms with an operating revenue of at least \$50M.
    \item \textbf{City network:} all board interlocks between firms with an operating revenue of at least \$5M, aggregated at the city level, similar to the network used in \cite{Heemskerk2016WhereDirectorates};
    \item \textbf{Country network:} all board interlocks between firms with an operating revenue of at least \$5M, aggregated at the country level, similar to the network used in \cite{heemskerk2016corporate};
\end{enumerate}

\noindent
From each network, we only considered the largest connected component (which in all cases captured over $95\%$ of all edges), and we do not consider self-loops (which correspond to the number of shared directors within one country or city). 
Basic descriptive statistics of these datasets are shown in  Table~\ref{tab:networks}. 
All networks are undirected, weighted networks. 
Next to the number of nodes and edges, we show the mean degree, indicating to how many other nodes a node is connected on average. 
Lastly, the mean weighted degree denotes the mean value of the weighted degree, i.e. the sum of weights of all edges adjacent to a node.

\subsection{Experimental setup} \label{sec:exp_setup}

For the threshold in the calculation of community membership consistency, as described in Section~\ref{sec:consistencymeasure}, we choose edge consistency threshold $\theta = 0.9$ based on the edge consistency distribution. \deleted{For example, in the country network, the fraction of edges with consistency value lower than 0.9 is 0.1, meaning we use the 10\% most inconsistent edges to define the community membership consistency for nodes. }
For the community cores and fringes, as described in Section~\ref{sec:communitycores}, we choose core and fringe threshold parameters of respectively \replaced{$\kappa = 0.9$}{$0.9$} and \replaced{$\phi = 0.5$}{0.5}.

Our method has been implemented in Python and is available as a python package \textit{nwtools}\footnote{\url{DOI: 10.5281/zenodo.3247681}}. Scripts and notebooks to generate the plots in this paper can be found online\footnote{\url{https://github.com/research-Dafne/consistency-paper}}.

\begin{table}[!ht]
\begin{center}
\caption{Descriptive statistics of the networks used. Density denotes the average number of edges per node.}
\label{tab:networks}
\begin{tabular}{|l||r|r|r|r|}
\hline 
Network & Nodes & Edges &  Mean degree & Mean weighted degree \\
\hline
\textbf{Countries} & 170 & 2,554 & 30.0 & 3312.4 \\
\hline 
\textbf{Cities} & 24,747 & 859,665 & 69.5 & 1238.3 \\
\hline 
\textbf{Firms} & 73,167 & 271,169 & 7.4 & 11.3 \\
\hline
\end{tabular}
\end{center}
\end{table}

\begin{figure}[!t]
    \centering
    \begin{subfigure}[b]{0.3\textwidth}
        \includegraphics[width=\textwidth]{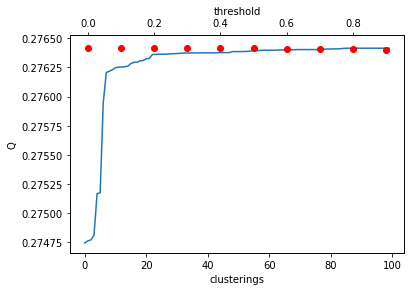}
        \caption{Countries}
        \label{fig:thresholds:countries1}
    \end{subfigure}
    \begin{subfigure}[b]{0.3\textwidth}
        \includegraphics[width=\textwidth]{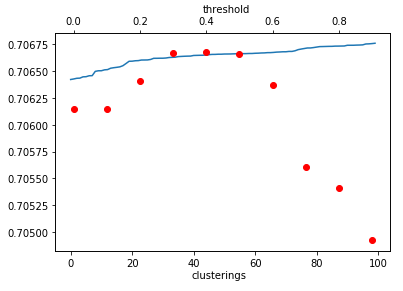}
        \caption{Cities}
        \label{fig:thresholds:cities1}
    \end{subfigure}
        \begin{subfigure}[b]{0.3\textwidth}
        \includegraphics[width=\textwidth]{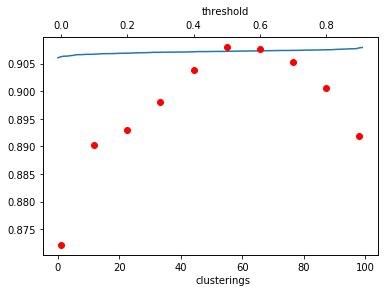}
        \caption{Firms}
        \label{fig:thresholds:firms1}
    \end{subfigure}
    \caption{The (sorted) modularity values of the individual clusterings (blue line) and of the consensus clustering for different thresholds (red dots)}
    \label{fig:thresholds}
\end{figure}

\begin{figure}[!b]
    \centering
    \begin{subfigure}[b]{0.3\textwidth}
        \includegraphics[width=\textwidth]{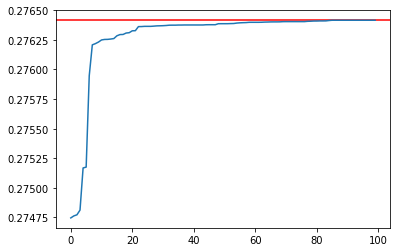}
        \caption{Countries}
        \label{fig:modularity:countries}
    \end{subfigure}
    \begin{subfigure}[b]{0.3\textwidth}
        \includegraphics[width=\textwidth]{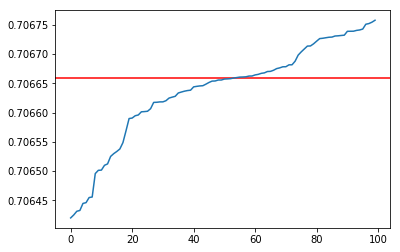}
        \caption{Cities}
        \label{fig:modularity:cities}
    \end{subfigure}
        \begin{subfigure}[b]{0.3\textwidth}
        \includegraphics[width=\textwidth]{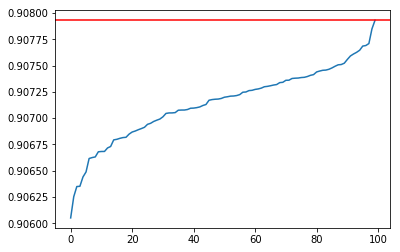}
        \caption{Firms}
        \label{fig:modularity:firms}
    \end{subfigure}
    \caption{The (sorted) modularity values of the individual clusterings (blue line) and of the consensus clustering for $\tau = 0.5$ (red line)}
    \label{fig:modularity}
\end{figure}

\subsection{Modularity landscape and consensus clustering} \label{sec:mod_landscape}
We describe the results from the consensus clustering, as described in Section~\ref{sec:consensus}, and investigate the degeneracy of the modularity landscapes for our networks, to show the necessity of a node community consistency score.

Figure \ref{fig:thresholds} shows the modularity value of the initial clusterings, obtained with the Leiden algorithm, for the different networks, together with the modularity values for consensus clustering with different thresholds. 
For all three networks, we proceed with threshold $\tau=0.5$ as this gives the highest modularity values. 
The resulting modularity value is plotted, together with the modularity values \replaced{of}{if} the initial clusterings, in Figure~\ref{fig:modularity}.
For all three networks, the range of modularity values of the different clusterings is very small, confirming the degeneracy in the modularity landscape. As can be expected, the consensus clustering does not always have the highest possible modularity value, but is similar to the modularity values of the original clusterings.

\begin{table}[!t]
\begin{center}
\caption{Results of modularity optimization for the three networks. We report the number of unique partitions, the mean modularity score of those partition, the modularity of the consensus clustering, the mean Normalized Mutual Information (NMI) of all pairs of clusterings, and the mean NMI between the consensus clustering and individual clusterings.}
\label{tab:nmi}
\begin{tabular}{|l||r|r|r|}
\hline 
Network & Countries & Cities & Firms
\\
\hline \hline Unique partitions & 64 & 100 & 100
\\
\hline Mean modularity & 0.276 & 0.707 & 0.907
\\
\hline Consensus modularity & 0.276 & 0.707 & 0.908
\\
\hline Mean NMI & 0.866 & 0.972 & 0.856
\\
\hline Mean Consensus NMI & 0.931 & 0.980 & 0.882
\\
\hline 
\end{tabular}
\end{center}
\end{table}

More details on the solutions of the modularity optimization are given in Table \ref{tab:nmi}, where we calculate the NMI (Normalized Mutual Information) between the consensus clustering and the individual clusterings, and among the individual clusterings. 
NMI is a similarity measure and in this context thus tells us how diverse the set of clusterings is in terms of similarity between clusterings. 
Although there are many different clusterings, the mean NMI  between clusterings is high for all three networks. 
The mean NMI between the consensus clustering and the individual clusterings is even higher. 
This suggests that the effects of the degeneracy for the overall clustering of these networks is not so pronounced, because the clusterings are all quite similar. This means we can safely take the consensus clustering as reference clustering when looking at community membership consistency in Section~\ref{sec:node_consistency}. 
However, the degeneracy can still affect individual nodes, as we will see later in the results. The large variety of high-quality clusters shows the need to obtain insights in the effects on the individual nodes.

\subsection{Community membership consistency} \label{sec:node_consistency}
\begin{figure}[!h]
\centering
\begin{subfigure}[b]{0.3\textwidth}
    \centering
  \includegraphics[width=\textwidth]{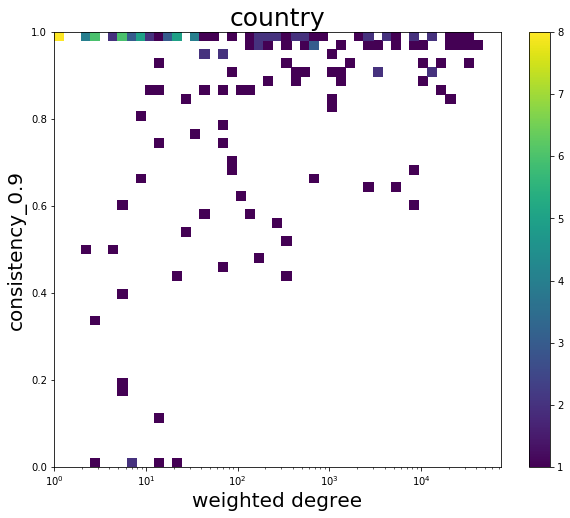}
  \centering
    \caption{Countries:\\ $\rho=-0.21, p=5.6\cdot 10^{-3}$} 
    \label{fig:nodeconsistency:countries}
\end{subfigure}
\begin{subfigure}[b]{0.3\textwidth}
  \includegraphics[width=\textwidth]{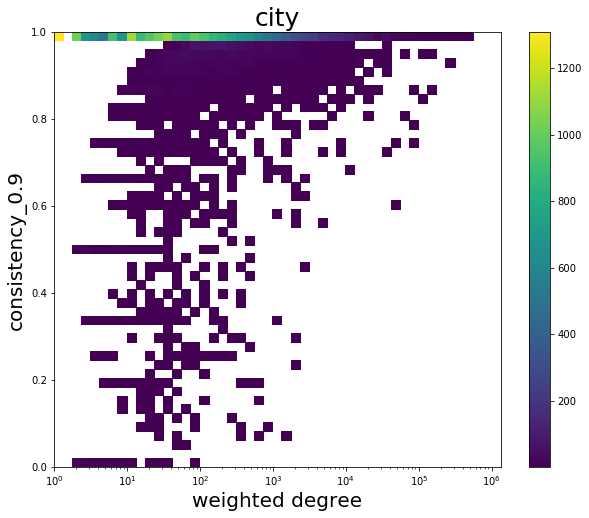}
    \caption{Cities:\\ $\rho=-0.36, p=0$}
    \label{fig:nodeconsistency:cities}
\end{subfigure}
\begin{subfigure}[b]{0.3\textwidth}
  \includegraphics[width=\textwidth]{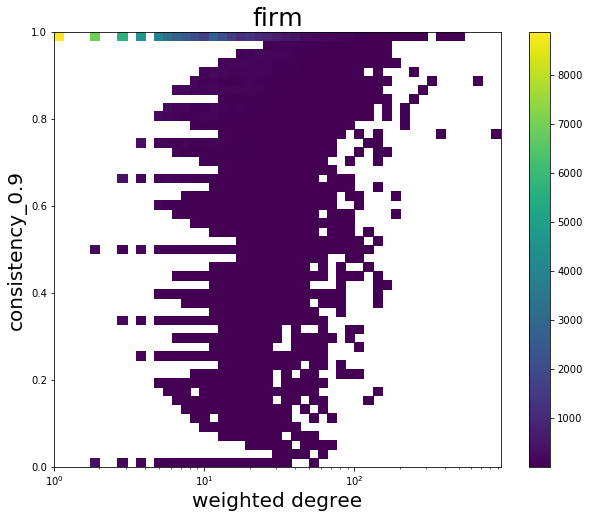}
    \caption{Firms: \\ $\rho=-0.15, p=0$}
    \label{fig:nodeconsistency:firms}
\end{subfigure}
\caption{Two-dimensional histogram of community membership consistency scores $s_i^{(0.9)}$ and weighted degree in the country, \replaced{city and firm networks}{network}, listing the Spearman correlation ($\rho$) between the two. Color denotes the number of nodes in the bin (cell).} 
\label{fig:nodeconsistency}      
\end{figure}

Here we investigate the node-level consistency scores, as proposed in Section~\ref{sec:consistencymeasure} and compare the values to other node-specific measures. 
Figure \ref{fig:nodeconsistency} plots the weighted degree of each node against its community membership consistency, and denotes the Spearman rank correlation between these values. These results suggest that the community membership consistency may be less informative for low-degree nodes, because there are fewer possible values of community membership consistency. In the most extreme case, nodes that have only one neighbor will have a community membership consistency of either 0 or 1.
An interesting observation is that community membership consistency is negatively correlated with weighted degree. \added{Note that based on the colors in the plot, there is indeed a high density of nodes in the upper left part bins, having a low weighted degree and high consistency score.}

In fact, the unweighted node degree also shows a negative Spearman correlation with community membership consistency (e.g. for the country network, $\rho=-0.38, p=0$). 
Note that, as can be seen in the plot, there are many low-degree nodes that have very high consistency. 
Thus, if a node has low community membership consistency, it is more likely to have a higher degree. 
High degree nodes are often viewed as interesting because of their central position in the network, but these results suggests we need to be careful in drawing conclusions about their positions in communities.

\subsection{Results}\label{sec:descriptive} 
\begin{figure}[!t]
  \includegraphics[width=\textwidth]{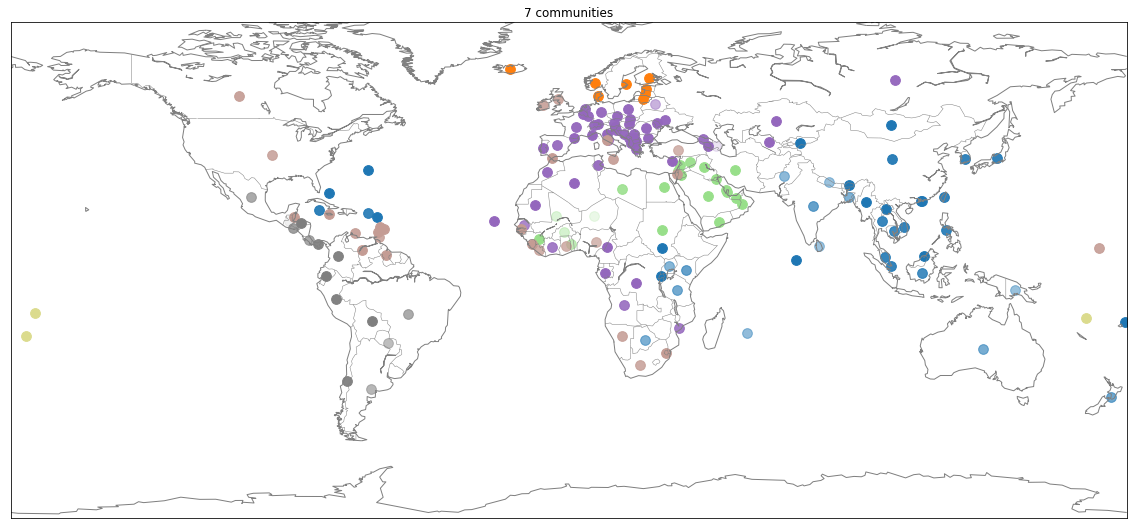}
\caption{Consensus communities for the countries. The transparency of the color denotes the community membership consistency $s_i^{(0.9)}$.}
\label{fig:map-countries}       
\end{figure}

We now turn to a descriptive analysis of the community detection outcomes, using the node consistency scores for meaningful interpretation. 
A geographic representation of the country network is presented in Figure~\ref{fig:map-countries}.
Here, the community membership consistency scores are visualized by means of the transparency of the corresponding node color. 
The color itself is based on the community assignment according to the consensus clustering. 
This allows us to easily identify the community of a node as well as the extent to which it is a consistent community member. 
Descriptive statistics on the composition of each community's hard \added{community} core, \added{community} core and fringe are shown in Table \ref{tab:countrycomms}. 
It also lists for each community the three heaviest nodes (nodes with the highest sum of weighted edges connected to it). 

\begin{table}[!b]
    \centering
    \begin{tabular}{|p{2.6cm}|p{0.8cm}|p{0.8cm}|p{0.8cm}|p{0.8cm}|p{0.8cm}|p{0.8cm}|p{0.8cm}|}
    \hline
Community &  I & II & III & IV & V & VI & VII
\\ \hline 
Number of countries  & 42 & 51 & 28 & 23 & 15 & 8 & 3
\\ \hline 
Mean consistency & 0.81 & 0.95 & 0.88 & 0.80 & 0.72 & 1.00 & 1
\\ \hline 
Hard \added{community} core members& 14 & 18 & 12 & 13 & 4 & 8 & 3
\\ \hline 
Hard \added{community} core (relative) & 33\% & 35\% & 43\% & 57\% & 27\% & 100\% & 100\%
\\ \hline 
\added{Community} core  & 24 & 45 & 17 & 16 & 7 & 8 & 3
\\ \hline 
\added{Community} core (relative) & 57\% & 88\% & 61\% & 70\% & 47\% & 100\% & 100\%
\\ \hline 
Fringe members & 5 & 1 & 1 & 5 & 2 & 0 & 0
\\ \hline 
Heaviest nodes & \makecell[l]{CN \\ SG \\ MY} & \makecell[l]{FR \\ IT \\ ES} & \makecell[l]{GB \\ US \\ IE} & \makecell[l]{AE \\ SA \\ EG} & \makecell[l]{BR \\ CO \\ CL} & \makecell[l]{SE \\ NO \\ FI} & \makecell[l]{WS \\ TO \\ VU}
\\ \hline 
    \end{tabular}
    \caption{Descriptives of the communities in the country network}
    \label{tab:countrycomms}
\end{table}

Community I, which is the second largest community, gravitates around Asia, with as its \added{community} core members China, Japan, South Korea, Singapore and Malaysia. Interestingly, this community also contains as core members well known offshore financial centres such as the Cayman Islands, and British Virgin Islands, corroborating previous work~\cite{takes2016centrality}. The \added{community} core has 24 members. This includes a sizeable number of relatively small Asian economies such as Bhutan, Mongolia, and Laos. 
There are five countries that score under 0.5 on consistency, including Mauritius, Papua New Guinea, Nepal and Sri Lanka. 
While both Nepal and Bhutan are members of this community, Nepal is a fringe member and Bhutan a \added{community} core member. 
The node consistency score helps us to see this important difference between these economies' membership of the Asian community. 
These findings corroborate previous work that finds a coherent Asian community in the board interlock network~\cite{heemskerk2016corporate}, but adds to this work by showing which smaller economies are fringe or core members. 

Community VI is of interest as it shows full consistency. The eight members of this Nordic-Baltic community are always positioned together. Earlier work already found that this cluster of countries is strongly interconnected~\cite{heemskerk2016corporate}. However, with the overview of consistency scores we can now conclude that all of the eight economies have a similar fit. The same goes for community VII, with as its only members three Pacific island states: Tonga, Vanuatu and Samoa. These three small economies are oriented upon themselves. 

Community V brings together the major economies on the South American continent, but in a relatively inconsistent manner. 
Under half of its members are part of the \added{community} core, and the large economic players such as Argentina ($0.56$), Mexico ($0.66$) and Brazil ($0.65$) are all outside the \added{community} core. 
This suggests low levels of economic integration through corporate board interlocks, corroborating previous findings on this matter (see for example \cite{cardenas2015latin}). 
A more detailed analysis would aim to reveal if the orientation of these core Latin economies are markedly different, for instance such that Mexico is more oriented to North America and Brazil to Europe.  

If we move on to community IV, we see 23 members, and a \added{community} core of 16 (hard \added{community} core of 13). 
It contains the key economies in the middle and near east, such as Saudi Arabia, Egypt, but also Iran and Lebanon. 
A few African economies are included in this community as well, but given their low consistency score we cannot consider Burkina Faso, Mali, or Niger meaningful members. 
Benin and Madagascar score zero consistency, which means that they are not integrated in this community at all. 
It is of interest to see economies that are strongly opposed in the geopolitical and military realm such as Iran and Saudi Arabia in one community together with Iraq and Syria. 
This suggest that the political and military divisions have not washed away the economic integration, although it should be noted that these connections may run through countries not involved in these conflicts. 
Interestingly, Sudan is placed with the Middle East cluster, whereas its geographical counterpart and military rival South Sudan is placed in the Asian community. 
Also, it is notably that the largest economy in the region, Israel, is part of another community, namely community IV.

Community III centres around North America (USA and Canada) and also contains Ireland, the United Kingdom, and South Africa. 
It reflects to some extent the old British Empire (see also~\cite{takes2016centrality}). 
However, the consistency scores hint at a rather interesting dynamic as the main economies UK (0.86) and the USA (0.84) in this community are not in the hard \added{community} core. 
Also, it is remarkable that Australia and New Zeeland are not in this community but rather placed as non-core members in community zero. 
Previous work found that the Anglophone cluster was a strong backbone of the transnational network of board interlocks. 
Our more detailed analysis shows that to the extent that this community is still discernible, it may be moving towards disintegration.  

This leaves community II, where the European economies are located. With 51 members it is the largest community we find. 
Given the high level of European economic integration and the relatively small geographic size of many of the European countries this may come as no surprise. Like other communities, we see that the hard \added{community} core members are typically smaller economies such as Moldova, Albania, and Kosovo. We also see that some small non-European economies such as Cameroon, Congo, and Algeria are firmly positioned in this community, signalling a European rather than an Asian or North American orientation. 
There are hardly fringe members in this community, and only Azerbaijan (with a consistency of 0.17) should not be considered as a member of this community from a substantive point of view. 
Of some interest is that The Netherlands is positioned in this European community, and not in the transatlantic Anglophone community as previous research found~\cite{heemskerk2016corporate}.

This descriptive analysis of the community detection results illustrates the usefulness and importance of considering node level consistency scores for a proper interpretation of the outcomes. 
An important observation is that economies with the highest consistency scores are typically smaller economies, while the larger and more dominant community members are typically in the \added{community} core with at least 0.9 consistency score.
This coincides with the negative correlation we found between weighted node degree (number of connections of a country) and consistency. 
We saw that almost all communities had some members with rather low consistency scores. 
This information allows us to refrain from any meaningful interpretation of these countries' particular results. 
The empirical outcome that Azerbaijan is positioned in the European community (consistency of $0.17$) or Guatamala in the Latin community (consistency of $0.11$) has no substantive meaning. 
This is an example of how considering the node consistency scores makes one prone to serious errors in the interpretation of the community detection results.

\section{Discussion} \label{sec:discussion}

The goal of this study was to offer a way to draw reliable conclusions from community detection. 
For this purpose, we introduced  community  membership consistency, based on the consensus matrix from different runs of a modularity maximization algorithm.  
We showed the value of this method for board interlock networks, where indeed there are central nodes for which we cannot rely on their placement in the community by the consensus clustering method.

It is important to note that modularity optimization is only one of many community detection methods available, and it has known limits, other than the degeneracy problem~\cite{fortunato2010community,Fortunato2007}. However, many other popular community detection algorithms, such as Infomap~\cite{rosvall2008maps} and OSLOM~\cite{Lancichinetti_Radicchi_Ramasco_Fortunato_2011}, are also non-deterministic or depend on node ordering, and may also result in different solutions upon multiple runs, essentially suffering from the same limitations.
Our proposed method is thus equally applicable to those algorithms.

As noted in Section~\ref{sec:introduction}, the proposed measure is similar to community (in)consistency as defined by~\cite{kim_community_2015,kim_relational_2019}, which uses the sum of squares of the distances to the consensus matrix for each node.
The definition used in that work differs from our definition in a few aspects. First, it computes the node-level score from the consistency values of all node pairs, in contrast to our approach where we only look at the direct neighborhood of a node. 
Second, it aggregates the node pair scores using the mean, somewhat mitigating the skewedness of the distribution of node pair consistency by defining it using a square function instead of absolute value.
Community inconsistency is then shown to be informative as a centrality metric. 
It thus serves a different goal than our community membership consistency, which is used to define \added{community} cores and fringe members and get a better understanding of node-specific community membership. 

The community membership consistency score heavily depends on the modularity landscape formed by the initial clusterings, which is shaped by the topology of the network. 
Explorations of the shape of the modularity landscape have been done in previous research.
For example, in \cite{Calatayud_Bernardo-Madrid_Neuman_Rojas_Rosvall_2019}, the solution landscape is investigated by clustering the resulting solutions. 
Other work \cite{de_santiago_role_2016} shows that low-degree nodes are most influential on the number of suboptimal partitions. 
This contradicts with our finding that community membership consistency is negatively correlated with (weighted) degree. 
 
Future work can investigate the relationship between community membership consistency and node properties such as degree. 
It can further explore the generalizibility of our method to different types of networks.  Note that the community membership consistency is highly dependent on the node degree, so it is possible that the method has limitations in very sparse networks. 
It may be interesting to study effects of the network structure on the interpretability of the results, as well as the choice of threshold $\theta$.

The extent to which the degeneracy leads to diversity in the clusterings, depends of course on the optimization algorithm, as was also shown by \cite{Good2010PerformanceContexts}. 
To calculate and use consistency in a sensible way, one could argue that a large diversity of clusterings is positive, as long as they are all close to the optimum. However, this is not what optimization algorithms are designed for and it is unclear to which extent different algorithms explore the modularity space. 
We have observed that the Leiden algorithm, which we use in this paper, results in a more diverse set of outcomes than its predecessor, the Louvain algorithm. 
In \cite{Riolo_Newman_2019}, a generative model is used so that it is possible to sample from the posterior distribution over clusterings. 
It would be interesting to combine this approach with our proposed consistency metric.

\added{Additionally, it should be noted that in the example of corporate networks, and in general for our approach, the multilayer aspect of  networks could be considered, taking into account the different types of links, in our case, e.g., both board interlocks and ownership links between firms}~\cite{takes2018multiplex}.  \added{However, when applying the proposed community membership consistency method to community detection methods for more complex network models, such as multilayer networks, the interpretation also becomes more challenging. It would make it for example difficult to do a systematic interpretation from a domain perspective as we did in Section~\ref{sec:results}. 
This warrants further investigation in future work. 
Even more generally, it could be investigated how our approach could be used in other complex networks models, such as temporal networks and higher-order networks}~\cite{lambiotte2019networks}.

Another direction of future research is the interpretability of nodes with low consistency and the relationship between consistency and community dynamics. When a node has low consistency, applied researchers may be interested into why the node is not a stable member of the community. 
\added{Of course this can be done from a domain perspective, attempting to see if there is substantive change going on in the system. 
Interestingly, unstable nodes could also be a sign that communities are simply not well separable. While this may suggest different things, one explanation may be that we are in fact dealing with overlapping communities. This could in particular be the case for high degree nodes. In future work it might be interesting to see to what extent unstable nodes are in fact members of overlapping communities as found by an overlapping community detection method}~\cite{xie2013overlapping}\added{. Yet another explanation for unstable nodes that may be valuable to investigate further, is} whether inconsistencies in the obtained clustering are a sign of instability in the network itself, and are thus  related to changes in the network structure over time. Investigating this further would obviously require dynamic network data.

\section{Conclusion} \label{sec:conclusion}

Community detection algorithms are widely used to understand the meso-scale structure of networks. 
This work contributed to quantitatively drawing micro-level conclusions about the community membership of individual nodes. 
To achieve this, we proposed community member consistency, which is a node-specific metric to indicate reliability alongside a division into communities using consensus clustering. 

We applied this metric on 
the global board interlock network, and showed that we can distinguish between \added{community} core members \deleted{of the community} and fringe members. We showed that the non-core members are sometimes high-degree nodes, and that the consistency metric prevents us from viewing these nodes as central in the community.
We found a negative correlation between community membership consistency and node degree, suggesting that it is more likely for a high-degree node to jump communities.

The proposed measure may prove useful in other social science contexts where community detection results are used to gain insight about the role of individual nodes within a community. 
Future work may investigate the generalizability of the method for different types of networks, and the relationship of consistency scores with higher order structural properties of the network. 
Finally, the metric could be used to reliably infer movements of nodes between communities in a dynamically evolving network.



\bibliographystyle{spphys}
\bibliography{references}

\end{document}